# Molecular Modeling of the Microstructure Evolution during the Carbonization of PAN-Based Carbon Fibers

Saaketh Desai, Chunyu Li, Tongtong Shen, Alejandro Strachan[*]

School of Materials Engineering and Birck Nanotechnology Center

Purdue University, West Lafayette, IN, USA 47906

## Abstract

Development of high strength carbon fibers requires an understanding of the relationship between the processing conditions, microstructure and resulting properties. We have developed a molecular model that combines kinetic Monte Carlo and molecular dynamics techniques to predict the microstructure evolution during the carbonization process of carbon fiber manufacturing. The model accurately predicts the cross-sectional microstructure of carbon fibers, predicting features such as graphitic sheets and hairpin structures that have been observed experimentally. We predict the transverse modulus of the resulting fibers and find that the modulus is slightly lower than experimental values, but is up to an order of magnitude lower than ideal graphite. We attribute this to the perfect longitudinal texture of our simulated structures, as well as the chain sliding mechanism that governs the deformation of the fibers, rather than the van der Waals interaction that governs the modulus for graphite. We also observe that high reaction rates result in porous structures that have lower moduli.

## 1. Introduction

Carbon fibers are increasingly the material of choice for many high performance composites, due to their high stiffness and strength, combined with their low density [1]. Early work in this area focused on carbon fiber (CF) manufacturing using pitch and cellulose (rayon) based precursors, reviews of which are provided in Ref. [1], [2]. However, Polyacrylonitrile (PAN) is currently the precursor of choice for high strength carbon fibers [3], [4]. There have been extensive experimental studies of the microstructure [5]–[8] of these fibers, all indicating that the fibers consist of long and aligned graphitic 'turbostratic' sheets, with defects like voids disrupting the continuity along the fiber axis. In the transverse cross-section, the carbon fibers were observed to consist of folded sheets, details dictated by the processing conditions and the type of initial precursor. This has led to many proposed schematics of the internal structure of these fibers [5], [7], [9]. Despite all this progress, while commercial carbon fibers can achieve a modulus similar to that of ideal graphite, even the latest high-strength fibers [10] achieve values lower than 10% the ideal graphite strength. The strength of the fibers is limited by the defects present in the microstructure, and the extent of the effect of these defects can be understood by a detailed study of the effect of processing conditions on the resulting microstructure and properties.

[*] Corresponding author: strachan@purdue.edu (Alejandro Strachan)

Recent advances in computational ability allow us to approach this problem from an atomic perspective. Recent work has focused on building some isolated structures like defective D-Loops [11], basic structural units (BSU) [12], or polycrystalline and multilayer graphite models [13], [14] and studying its effect on the strength. While this work suggest some possible failure mechanisms, the predicted strength values are an order of magnitude greater than experimental values. There has also been work on identifying the mechanisms of the reactions occurring during carbonization [15], attempting to provide an atomic picture of the reactions leading to the formation of graphitic sheets. Despite this progress, no model to date can predict the CF microstructure starting from the stabilized CF structure, and predict the mechanical properties corresponding to these microstructures. The objective of this work is thus to take a first step towards developing a model that predicts the microstructure evolution during the carbonization process, with the aim of developing a framework that can quantitatively relate the processing variables (like reaction rate and temperature) to the generated structural features (like the length of graphitic sheets), and hence, the properties. The model employs a kinetic Monte Carlo (kMC) scheme to describe the chemical reactions occurring the carbonization, while using molecular dynamics (MD) to describe the relaxation of the system during the crosslinking process.

# 2. Molecular model of carbonization/graphitization

## 2.1 Scope and overview

The conversion of PAN precursor fibers to carbon fibers involves three major steps: stabilization, carbonization and graphitization, each of which involves multiple chemical reactions. A review of these manufacturing processes can be found in Refs. [2], [16]. Stabilization involves heating in air at a temperature of 200-300°C under which conditions a series of chemical reactions transform the PAN chains into structures that can withstand the high-temperatures required for carbonization without decomposing. While several reactions are believed to take place during stabilization, there is consensus that the result is the conversion of PAN chains into ladder like structures [17]. In the carbonization stage, the stabilized ladder like structures are heated to a temperature of 1000-1700°C [4], converting ladder like structures into the eventual carbon fiber microstructure, consisting of graphitic sheets. Lastly, in the graphitization stage, the fibers are heated up to 2500-3000°C to obtain high modulus fibers [16].

A reactive MD study using the ReaxFF force field [15] has provided an atomic picture of the first steps of carbonization and suggested the elimination of gases like $N_2$, $H_2$, $NH_3$ and HCN, along with cyclization reactions, leading to formation of five membered rings and eventually six and seven membered rings. These predictions are consistent with experimental observations. However, the use of reactive MD severely limits the timescales accessible and precludes the study of microstructure evolution. Our motivation to study microstructure evolution and predict the final CF microstructure dictates our use of a coarse-grained approach and combined kMC and MD. In this endeavor, we ignore the specific details of the reactions and chemical environments occurring during carbonization and graphitization. Instead, we describe a generic, averaged version of the individual processes, and describe carbonization and graphitization as chemical reactions between carbon atoms in nearby ladder structures to create $sp^2$ bonds leading to graphitic sheets.

## 2.2 Carbonization/graphitization method

We start the process with a well-relaxed simulation cell containing a number of coarse grain ladder structures representing the stabilized fiber. Our coarse grained approach ignores the details of various chemical reactions responsible for graphitization [2], [17] and the initial ladder structure, see Figure 1(a), thus ignores the specific chemical nature of the ladder structures such as heteroatoms and side groups. The ladder structure consists of two types of carbon atoms: saturated sp$^2$ carbon atoms bonded to three other carbon atoms, marked as C in Figure 1(a), and reactive atoms, marked as C*. Each chain is infinitely long (with 8 atoms per unit cell) and perfectly aligned along the Z direction of the simulation cell, see Figures 1(b) and (c). We stress that assuming that the chains are perfectly aligned is an approximation and that in this first effort to model microstructure evolution, we are interested in predicting the cross-sectional CF microstructure and properties. A set of these chains are packed into a simulation cell with periodic boundary conditions in all directions. Before crosslinking, this initial structure is relaxed and the details of this procedure are described in sub-section 3.1.

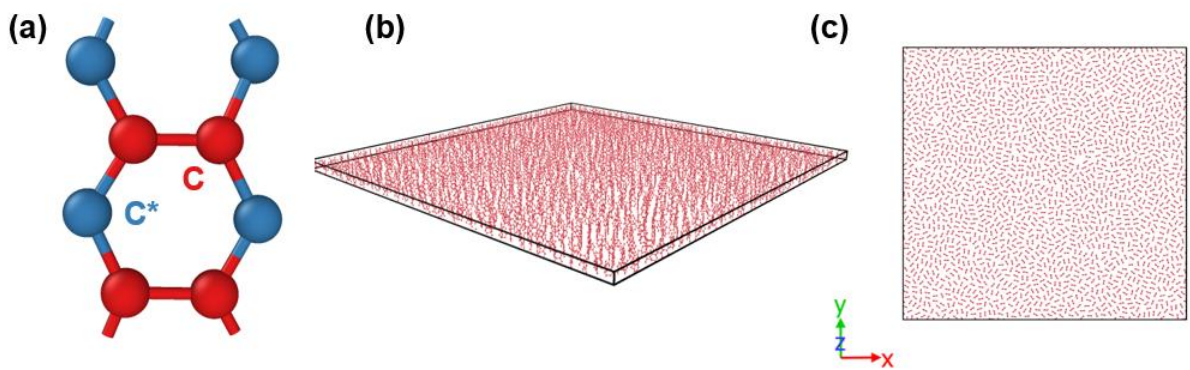

*Figure 1: (a) The initial monomer configuration, where red atoms indicate the saturated sp$^2$ carbon atoms and blue atoms indicate the 'reactive atoms', with only 2 bonds (b) Perspective view of the packed monomers. Note the small out-of-plane thickness of the simulation cell (c) Top view of a representative relaxed structure*

The graphitization model involves cycles of bond creation followed by relaxation using MD, see Figure 2. A similar procedure has been applied previously to study crosslinking in epoxy-amine systems [18], [19] and has been successful in predicting a wide range of properties such as modulus and yield strength. The key inputs to the carbonization/graphitization model are the molecular structure of the initial ladder structures and the rate of the reactions (bond formation). In reality, carbonization would involve multiple reactions like formation of polycyclic chains and the evolution of various gases, and the rates of each of these processes would affect the reactivity of the chains and hence, the final structure. The rate of these reactions will depend on the activation energy associated with this process, and a pre-factor that depends on the entropy of the reactants and transition state [20]. This activation energy can be obtained using transition state theory and electronic structure calculations, as is the custom in kMC simulations. The activation energy will depend on the local chemical environment (ignored in our reaction model), the separation distance and the relative orientation of the molecules.

In our model, we acknowledge that atoms further away are less likely to react, and thus impose a distance cutoff to identify possible reactive pairs, see Step 1 in Figure 2. Given this set of nearby C* atoms, the energetics of bond formation, and consequently reaction rates, will be affected by the relative orientation between the two molecules. Thus we impose an angle cutoff, see Step 2 in Figure 2. The orientation of two molecules is described by the angle between the bond joining the two reactive atoms and the plane of the molecule associated to the reactive atoms. Thus, given two reactive atoms, there are two improper angles possible and the atoms are eligible for bond creation only if both these angles are within the imposed cutoff.

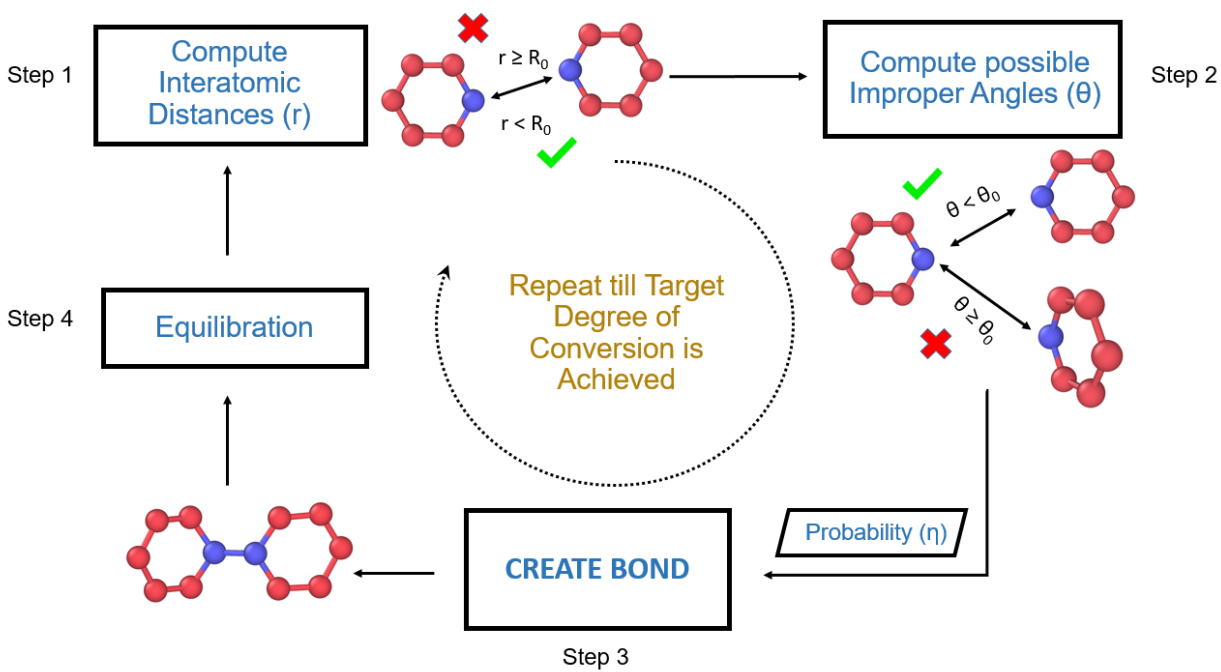

*Figure 2: Schematic of the crosslinking algorithm employed, where the blue atoms are the ones considered for bond creation. Here, '$R_0$' indicates the distance cutoff used and '$\theta_0$' indicates the improper angle cutoff. The probability '$\eta$' can range from 0 to 1.*

The dependence of reaction rate on separation distance merits additional discussion. We find that there are two characteristic distances that are important for graphitization, see Figure 3 (a). Nearest neighboring ladder chains with no covalent bonds between them are separated by typical van der Waals distances of approximately 3.5 Å. However, once a bond is created between these chains, the separation of nearby reactive carbon atoms is reduced to less than 3Å, see Figure 3 (b). Clearly, these reactive atoms in close proximity will have a higher reaction rate than those at van der Waals distances and this is taken into account in our method. The importance of using a two cutoff model and an angle constraint is further described in section 4.1.

In this first effort to model the graphitization of the ladder chain structures, we define a simple set of rules to determine chemical reactions and study how they affect the resulting microstructures and properties. Given all pairs of reactive atoms that fall within the capture distance ($R_0$) and whose ladder chains have an angle mismatch less than a threshold variable ($\theta_0$) we select the possible reactions as pairs of atoms $i$ and $j$ such that such that $j$ is the closest reactive atom to $i$ and $i$ is the closest reactive atom to $j$. All these possible reactions are considered to have equal reaction rates. At each bond formation cycle a pre-determined fraction ($\eta$) of these reactive pairs are bonded, chosen stochastically. Following each cycle of bond creation with $R_0$ and relaxation we reduce the capture radius to 2.85 Å and perform three cycles of bond creation and relaxation with $\eta=1$ to account for higher rate of reaction for pairs of atoms with shorter separation distances shown in Fig. 3(b).

**Simulation time**: The simulation time has contributions from both the kMC bond creation events and from the MD relaxation. Within the kMC formalism, after selecting a process with rate $k_i$, the time should be advanced by picking a random number corresponding to an exponential probability distribution, characterized by the sum of all the reaction rates ($k_i$) of the system. Thus, the time in each bond creation cycle involving $n$ reactions is the sum of $n$ stochastic numbers obtained by $n$ samplings of this probability distribution. Each cycle of bond creation is followed by an MD relaxation of time 20 ps. Note that this is significantly shorter that the kMC time as the structure relaxation (following bond creation) occurs relatively fast due to the high stiffness of the graphitic sheets.

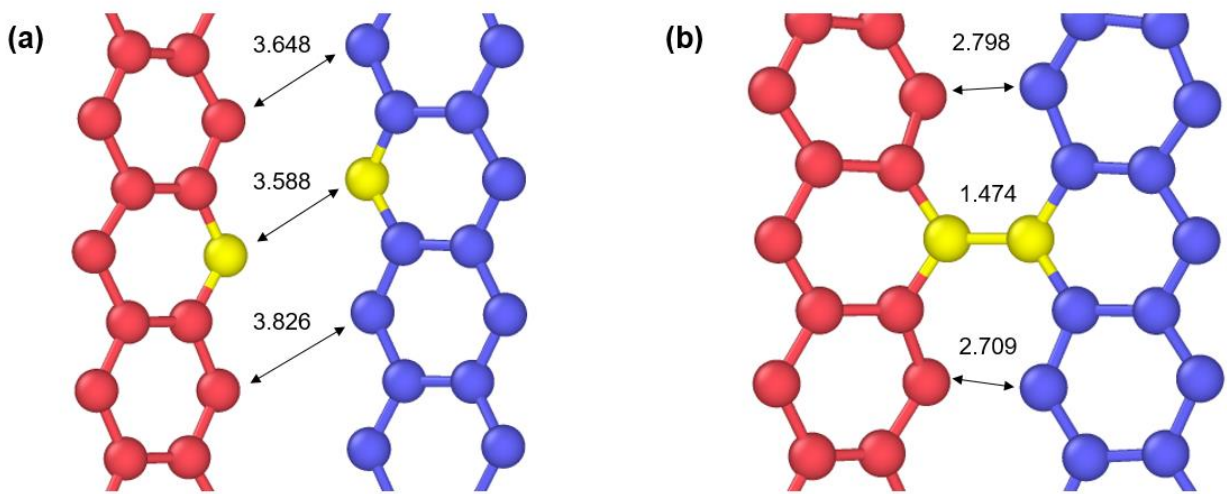

*Figure 3: (a) Simulation snapshot showing the change in chain structure after bond creation between two representative atoms (marked yellow). The atoms surrounding the bonds are observed to be close, allowing for a two cutoff model (numbers indicate distances in Å).*

## 3. Simulation details

### 3.1 Initial structure relaxation

All simulations are performed using the LAMMPS software package [21] and the atomic interactions are described by the DREIDING force field [22]. We use a Lennard Jones form to describe the non-bonded (van der Waals) interactions. Both reactive and saturated atoms are treated as $sp^2$ carbon atoms, using the default DREIDING parameters. The time step used to integrate the equations was 1 fs unless otherwise specified. The temperature is controlled by the Nosé-Hoover [23], [24] thermostat, with a damping constant of 0.1 ps. Similarly, the pressure is controlled by the Hoover barostat [25], [26], with a damping constant of 1 ps.

The simulation begins by creating a ladder-like chain monomer, as shown in Figure 1(a). These infinite chains are packed into the simulation cell with random orientations in the XY plane, at a density of ~ 0.5 g/cc, as shown in Figure 1(b). After packing, we relax the structure via energy minimization, using the conjugate gradient method with an energy tolerance of $10^{-6}$. The system is then relaxed at constant volume and at a temperature of 300 K (NVT ensemble) for 50 ps. We then equilibrate the system under at constant pressure and temperature (NPT), at atmospheric pressure, till a constant density of 1.38 g/cc is achieved, this step requiring 1.5 ns. We couple cell parameters along the X and Y (in plane) directions in the barostat to retain a square cross-section.

The next step is to take the system to the temperature at which carbonization/graphitization will be modeled, which is chosen to be 2500K to represent experimental conditions [16]. This is done in multiple steps to ensure a well-relaxed structure. We begin by heating the relaxed structure from 300K to 2500K at NVT conditions at a rate of 10K/ps. The system is then equilibrated under NPT conditions with a compressive stress of 0.5 GPa in the transverse directions to ensure a good packing of the chains, until the density achieves steady state (1 ns). This stress is then relaxed to 1 atm in 100 ps and the system is finally relaxed under NPT conditions, at 1 atm for (4 ns), enough to fully equilibrate the system. Throughout the procedure, the barostat maintains a square cross-section of the simulation cell.

Given that our crosslinking algorithm is stochastic in nature, we generate statistically independent samples to employ the crosslinking. This is done by using the relaxed structure from above and continuing to relax it under NPT conditions for 120 ps. From this trajectory, we pick 6 samples, each 20 ps apart.

## 3.2 Carbonization and graphitization

These samples are then crosslinked using the scheme described in Section 2. The method was implemented as an extension to the existing LAMMPS fix bond/create command and the code is available as supplementary material. During the crosslinking, the time step is lowered to 0.25 fs to avoid large atomic displacements after bond creation, as was observed in Ref. [18]. After each bond creation cycle, an energy minimization is performed, using the conjugate gradient method with an energy tolerance of $10^{-6}$. During the minimization, the atoms are only allowed to move 0.05 Å per step, allowing for a gradual descent in the energy of the system. After the minimization, the system is relaxed for 20 ps under NPT conditions at atmospheric pressure and the graphitization temperature.

## 3.3 Evaluating properties

Given that the simulation cell thickness is only ~5 A° during the crosslinking (as shown in Figure 1(b)), we first replicate the carbon fiber structures in the fiber (Z) direction to ensure a box length greater than twice the force field cutoff. This allows us to predict the properties accurately. The structures are then cooled down to 300K (under NPT conditions at rate of 10K/ps) before relaxing at 300K for 200 ps, also under NPT conditions, at atmospheric pressure. At this stage, we uncouple the X and Y simulation cell parameters in the barostat to avoid residual strains. To evaluate the transverse moduli, the relaxed structures from above were strained up to 5% in the X and Y direction, at a rate of $5 \times 10^9$ s$^{-1}$. A linear fit to the stress-strain curve gives the young's modulus in each direction.

## 4. Microstructures obtained during carbonization/graphitization

### 4.1 Role of multiple cutoff distances and angle constraints

Before presenting a systematic study of how the parameters in the graphitization model affect microstructure and properties, we discuss the importance of using a two cutoff model and accounting for the high reaction rates of reactive atoms at short distances (due to a nearby bond connecting two ladder chains). Similarly, we examine the effect of including a torsional angle constraint on the resulting microstructure.

Figure 4 shows a series of structures obtained at 300K, employing different choices for the distance and angle cutoffs. For 4 (a), we only use a single distance cutoff ($R_0$) of 5Å, while (b) and (c) employ an angle constraint ($\theta_0$) of 60°. Figure 4 (c) additionally employs the two cutoff model, with $R_0 = 5$Å and $R_1 = 2.85$Å. In all cases, the probability ($\eta$) is 0.1.

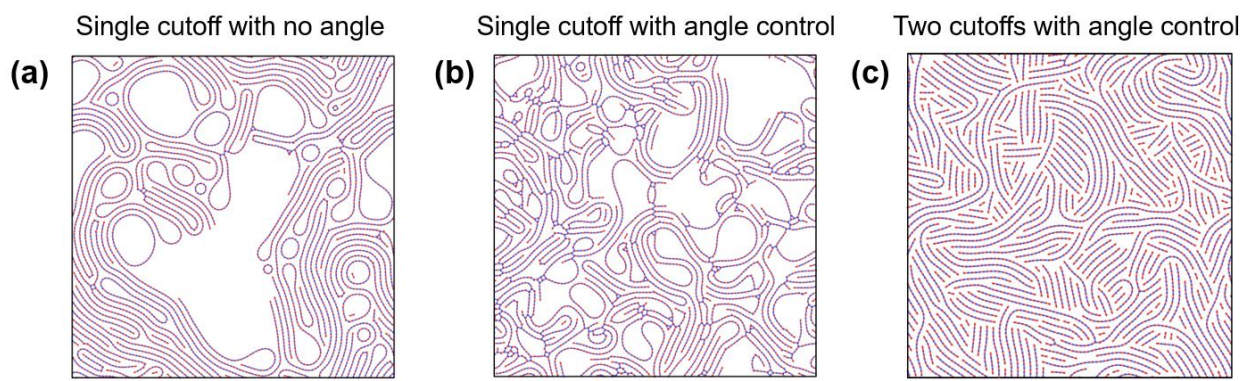

*Figure 4 (a) Structure obtained using a single cutoff has unrealistic nanotube structures (b) Implementing an additional angle control results in a porous and branched structure (c) Using the two cutoff model results in structures similar to experimental PAN based fibers*

Figure 4 (a) highlights the importance of angle constraints. The lack of an angle restriction enables bond formation between poorly oriented molecules resulting in a large number of loops (nanotubes) and open structures. A bond between poorly aligned ladder chains has a high activation barrier and consequently very low rate. Using a single cutoff distance with angle control also results in a disordered structure that does not contain the long graphitic sheets, see Figure 4(b). This is due to the fact that the initial monomer has two sets of reactive atoms and using a single cutoff can result in one atom bonding with one chain, while the other reactive atom of the same molecule might bond with another chain. Finally, Figure 4 (c) shows that the combination of two-cutoff model with angle control which results in structures with high packing density with microstructures similar to experimental PAN based fibers; a more quantitative comparison will be presented below.

## 4.2 Microstructure validation

Figures 5 (a-b) compare our predicted microstructure for $R_0=5$Å, $R_1=2.85$Å, $\theta_0=60°$. and $\eta=0.1$ with an experimental HRTEM image corresponding to high strength and high modulus gel spun fibers PAN copolymer fibers [10]. We find that the simulated structure contains key microstructural features like hairpins and curved graphitic sheets, similar to those observed experimentally.

To further validate our structures, we use LAMMPS to simulate a Wide Angle X-Ray Diffraction (WAXD) pattern (details in Supplementary), as shown in Figure 5 (c) and (d) and compare it with an experimental measurement on high modulus, low strength, PAN based GY-70 fibers, made by BASF [27]. In Figure 5 (c), we use the indexing notation followed in [27], where the (100) planes are stacked in the zigzag direction in the basal plane of the graphite sheet, while the (110) plane are stacked in the armchair direction. As seen in Figure 1 (a), initial ladder structure is oriented such that successive reactions will extend the graphitic sheet in the armchair direction, adding (110) planes. The relative intensity of the (110) and (100) planes depends on sample size due to the high degree of orientation and we, thus, scale intensities to match the (110) peaks between theory and experiments. The key features are the width of the (110) peak and the significant broadening of the (112) peak which the simulations capture. The width of the (110) peak is slightly underestimated in the predicted structure; this implies longer graphitic sheets than the specific carbon fiber characterized in the experiment.

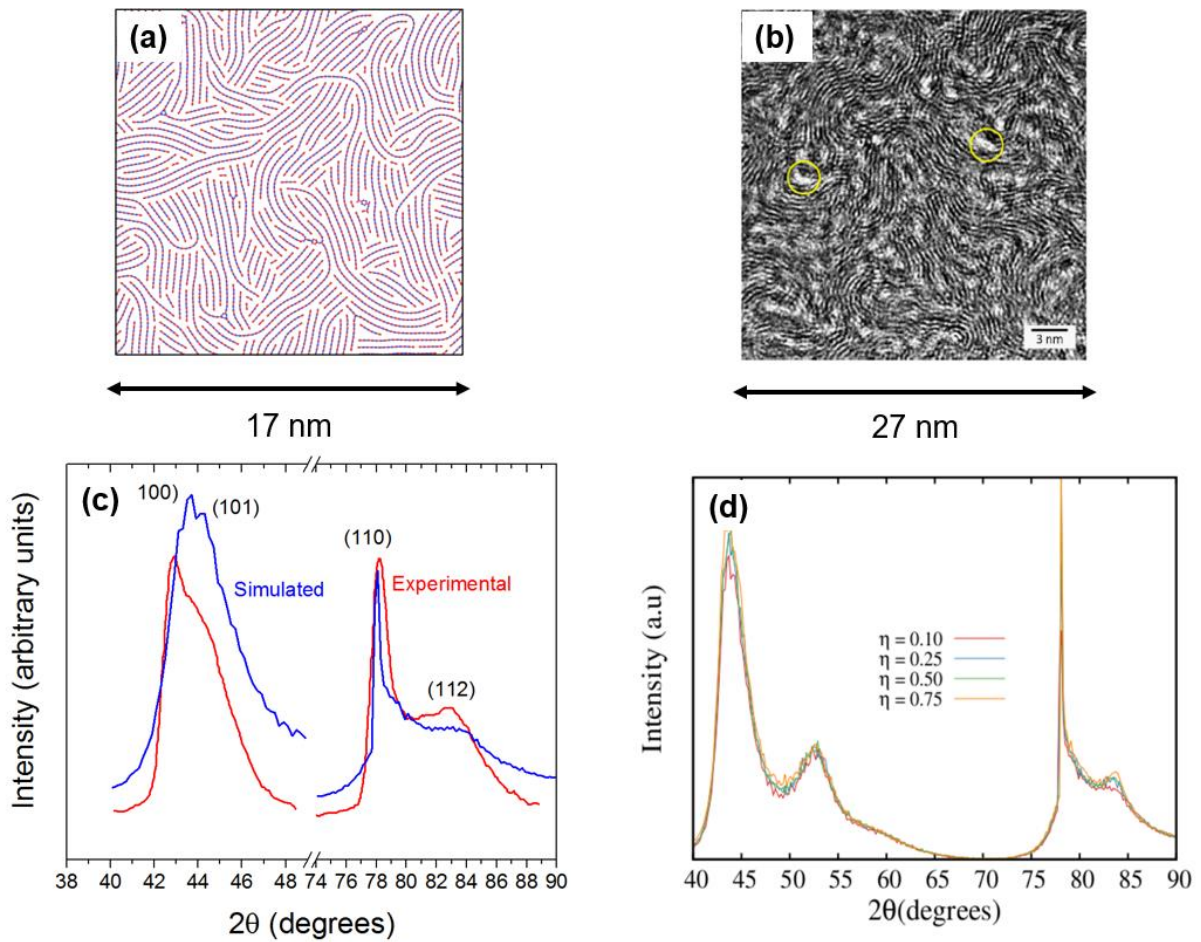

*Figure 5 (a) Top view of a representative simulated structure (b) An HRTEM image of a carbon fiber cross section, taken from Kumar et al, Carbon (2015) Reproduced with permission (c) The blue curve shows a simulated powder XRD pattern, while the red curve shows an experimentally observed XRD pattern, taken from Kumar, Anderson et al, Journal of materials science (1993).. The (101) and (112) peaks indicate extent of 3D order in the carbon fiber, an aspect that the extended 3D model will attempt to capture. (d) XRD patterns averaged over all six samples, for various reaction rates*

From the simulated peak width, we obtain an equivalent crystallite size of ~18 nm using the Scherrer equation. Typical PAN fiber crystallite size, denoted as $L_a$, ranges from 5 to 8 nm [1] although larger values have been reported. This overestimation of the crystallite size is expected, since we assumed infinitely periodic and perfectly aligned chains, resulting in a strong texture. In reality, longer chains would deviate from the fiber axis by about 15-25° [1], reducing the strong preferred orientation and decreasing the peak intensity. The peak at ~43.5° is a mixture of the (100) and (101) peak and may be attributed to the replication process employed in this study, which assumes greater order in three dimensions than might be revealed by using a full 3D model, beginning with an initial structure consisting of longer chains. We do not report the (002) peak as the 2D nature of the model ensures that the graphite planes in the fiber axis direction will be spaced at the equilibrium van Der Waals separation. The (002) interplanar spacing will be of prime focus in the full 3D model as it determines the young's modulus along the fiber axis. Figure 5 (d) shows that the XRD pattern varies little for various bond creation rates, indicating the structure of the graphitic sheets, characterized by the bonded C atoms and their van der Waals separation, remains similar, with variations occurring only in density, as will be shown in Section 5.

## 5. Process of curing and predicted properties

### 5.1 Evolution during carbonization/graphitization

Figure 6 shows the microstructure evolution during the crosslinking process for a representative sample. The parameters used for the simulation shown are $R_0=5$Å, $R_1=2.85$Å, $\theta_0=60°$ and $\eta=0.1$. The snapshots show the process by which the ladder chains crosslink and form graphitic sheets that grow in length with increasing conversion. A consistent feature of this process is the volume shrinkage, occurring due to the fact that unsaturated atoms that were previously at a van der Waals separation (~3.5 Å), are brought together to ~1.42 Å (the equilibrium $sp^2$ bond distance).

Figure 7 (a) shows the evolution of cure degree, as a function of MD simulation time (ignoring the time associated with the kMC steps) for various bond creation rates (represented by the probability $\eta$). The degree of conversion is defined as the ratio of the number of bonds created to the total possible number of bonds that can be created. A smaller $\eta$ represents a smaller number of reactions per kMC cycle and consequently shorter kMC time, i.e. a slower conversion rate, see Figure 7(b). We find that conversion degrees close to 90% can be achieved except for the fastest conversion rates where the MD simulation time is not long enough to enable full relaxation. The total MD relaxation times vary between 200 ps and 1 ns, these are comparable to the scales used to crosslink polymers with atomistic simulations [18], [28]. At the start of the crosslinking, we see that the number of reactions is directly proportional to the reaction rate, with a high probability corresponding to a high number of reactions, see Figure 7(a). As carbonization/graphitization occurs the number of reactive carbon atoms decreases and so does the number of reactions.

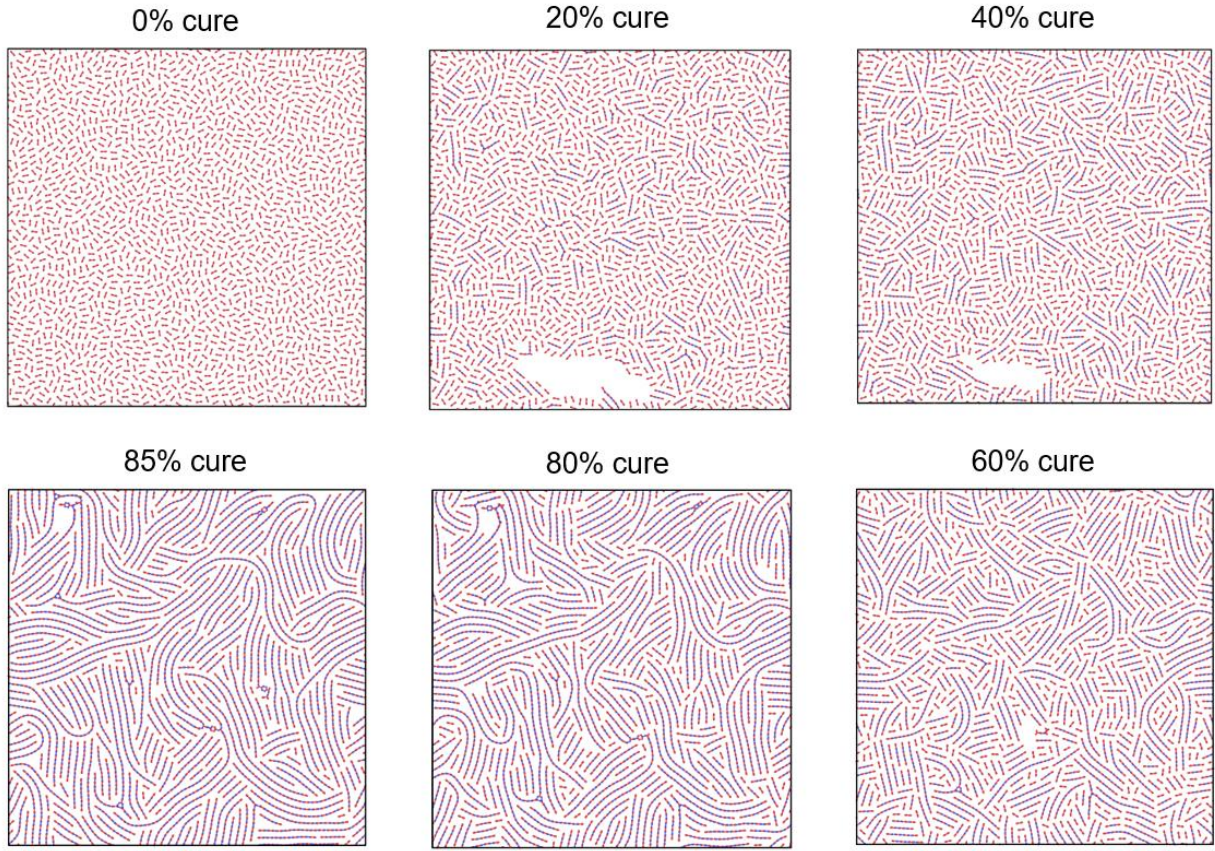

*Figure 6 Top view of the microstructure evolution during a sample crosslinking process. The parameters used for this process were*: $R_0=5Å$, $R_1=2.85Å$, $\theta_0=60°$ and $\eta=0.1$

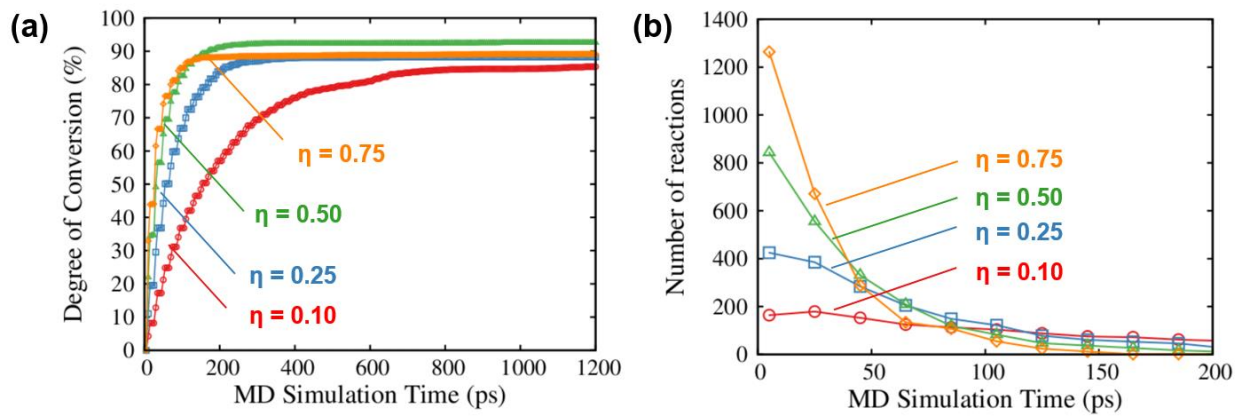

*Figure 7 (a) Time evolution of the degree of conversion for various probabilities (b) Time evolution of the number of reactions, for various probabilities, shown for the first 200 ps (MD time).*

Figure 8 (a) shows the evolution of density with the degree of conversion, for various probabilities ($\eta$). We find that the density increases with the conversion degree, again indicative of the fact that as more bonds are created, a greater number of unsaturated atoms move from a van der Waals separation to the equilibrium $sp^2$ bond distance. However, this trend is observed only until ~60% conversion, with the higher bond creation rates showing a subsequent drop in density. A large bond creation rate results in many bonds being created in the immediate neighborhood of an atom at each step. This results in graphitic sheets that are unable to relax and pack efficiently, resulting in excluded volumes that remain as voids throughout the rest of the simulation. The shaded area in Fig. 8(a) represent the typical range of experimental values for the density of PAN based carbon fibers [1]. Our predicted structures overestimate the density by approximately 10%. We attribute this observation the fact that we use infinitely periodic, perfectly aligned chains. This results in an unrealistically high degree of ordering in the fiber direction.

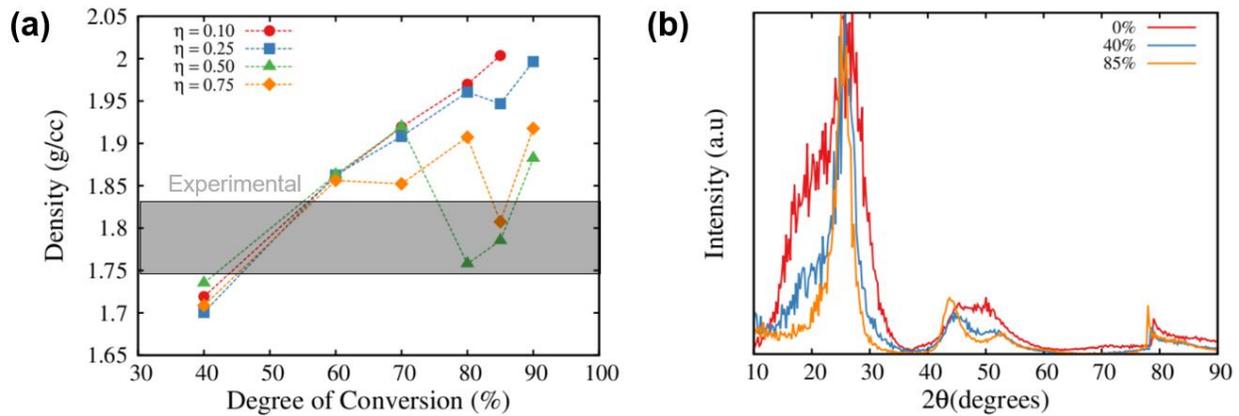

*Figure 8 (a) Evolution of density with degree of conversion (b) Evolution of the simulated XRD pattern with degree of conversion*

Figure 8 (b) shows the evolution of simulated XRD patterns with time. We observe that the (002) peak, corresponding to the van der Waals separation in graphite, decreases in width, denoting the evolution of a crystalline graphitic structure as the simulation progresses. Also, the shoulder initially present between 40 and 60° reduces to a peak at ~43.5°, corresponding to the (100) plane.

## 5.2 Properties of final carbon fibers

Figure 9 shows the average transverse modulus as a function of density for fully converted fibers created with various bond creation probabilities ($\eta$), capture radii ($R_0$) and angle control thresholds ($\theta_0$). The predicted moduli range from ~1.5-4 GPa. Before discussing how the model parameters affect the predictions we discuss our stiffness values.

It is useful to compare our predictions with the transverse modulus of graphite, which is reported to be around 36 GPa [29]. The MD trajectories during uniaxial tension of the converted fibers show that the prominent mode of CF deformation is via sliding of the chains across each other, whereas the modulus of graphite is a measure of the stiffness of the van der Waals attraction between the graphite layers.

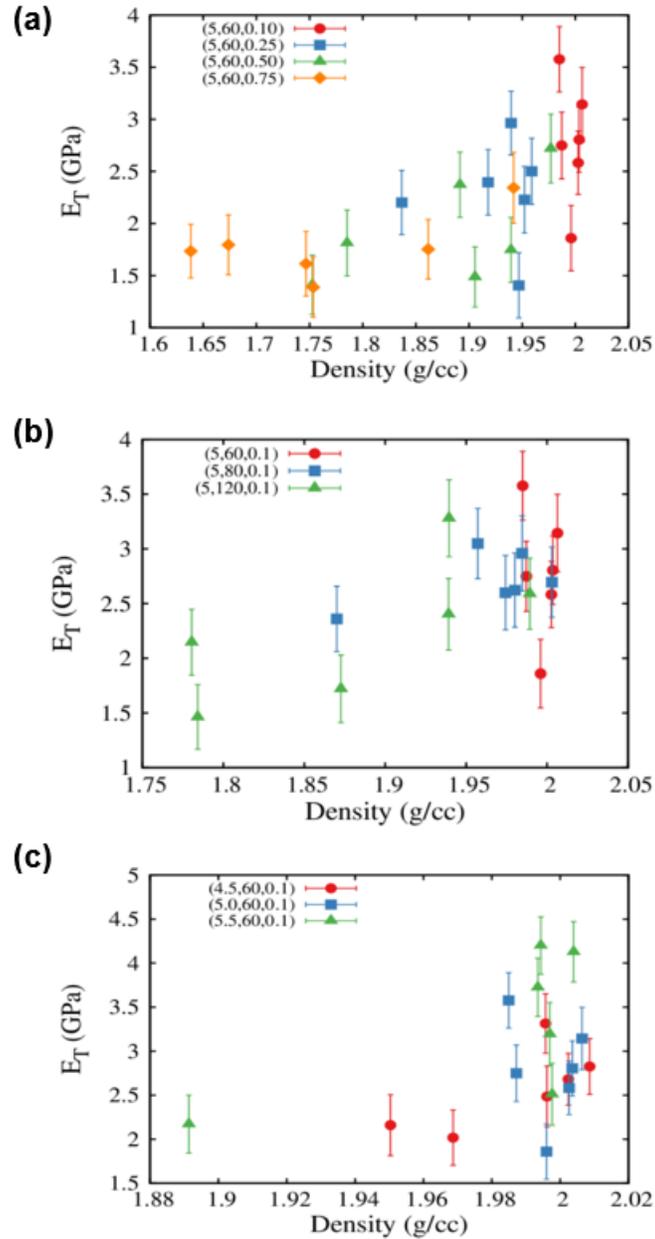

Figure 9. For each panel, the legend is a set of three numbers (l,m,n) where 'l' represents the distance cutoff (in Å), 'm' represents the angle cutoff (in degrees), and 'n' represents the

*probability. For each set of parameters, the individual points represent the 6 samples, all at 85% conversion.*

Given that the shear modulus for ideal graphite is ~4 GPa [30], it is clear that chain sliding is a low activation barrier process compared to increasing the van der Waals separation, and can thus occur at lower stresses, explaining the order-of-magnitude difference in the moduli, even after correcting for the different densities.

The CF transverse modulus has been experimentally estimated using single fiber compression tests resulting in values in the 6-10 GPa range [31], as well as nanoindentation [32] measurements that yield values of 9-15 GPa. Both experimental values are higher that the predicted average. In case of the single fiber compression test, the modulus is predicted by fitting the experimental data to an analytical equation relating the change in fiber diameter to the applied load using anisotropic elasticity [33]. In the case of the nanoindentation measurements, the modulus is extracted from the load displacement curve by using the Oliver-Pharr method [34], which assumes that the material is isotropic. However, carbon fibers are very anisotropic in nature, with a stiff longitudinal axis and a compliant transverse axis. The work of Vlassak [35] indicates that the indentation modulus for an anisotropic material will be some weighted average of the moduli in the various orientations, including the very stiff longitudinal direction. Thus, the nanoindentation values reported represent an upper bound to the fiber moduli [32]. Since the force field we use describes the stiffness tensor of graphite accurately, we attribute our underestimation to the fact that the chains are perfectly aligned. Misoriented crystallites, $sp^3$ bonds between graphitic sheets and amorphous regions in the experimental fibers will result in higher transverse moduli.

We now discuss the effect of model parameters on the predicted properties. Figure 9 (a) shows that increasing conversion rate results in a decrease in transverse modulus and density. This is because reducing the effective MD simulation time precludes the graphitic sheets to fully relax, resulting in significant internal strain and the formation of voids. Similarly, employing a loose angle constraint, see Figure 9 (b), results in folded sheets and nanotube-like structures that enclose a volume that cannot be occupied by other chains, resulting in poor chain packing and lower densities and, consequently, lower moduli. Varying the capture radius ($R_0$) (or distance cutoff) from 4.5 to 5.5 Å does not result in significant changes in density or stiffness, see Figure 9 (c). Interestingly, comparing predictions across all parameter sweeps, we observe that structures with nearly identical densities can have moduli varying by a factor of two. This is indicative of the large fluctuations expected in relatively small systems with complex microstructures. Further analysis should be performed to identify particular arrangements and lengths of the chains that result in easier sliding and thus, lower modulus.

## 6. Conclusions

We introduce the first model to describe the processing of carbon fibers. The model considers the carbonization and graphitization of coarse grained ladder chain structures and results in microstructures with the key structural features observed in experiments. The main inputs to our model are: (i) the initial molecular structure determined by the arrangement of the ladder chains that represent the molecular structure of the stabilized carbon fiber, and (ii) the parameters used to determine the rates of the bond formation processes that convert ladder chains to graphitic sheets. Regarding the first item, in this first effort we assume the chains to be perfectly aligned and infinitely periodic along the chain axis. This results in smaller simulation cells and enabled us to explore several aspects of the model effectively. The limitation is that we only predict the transverse microstructure and properties. Ongoing work is exploring larger simulation cells where long, finite chains are packed and crosslinked. In order to determine rates for chemical reactions we use a simple but physically–based approach based on the separation distance between carbon atoms and the relative angle between the ladder chains.

The model predicts key microstructural features known to exist in carbon fibers and the predicted diffraction patterns are in good agreement with experiments. The predicted densities are approximately 10% higher than experimental values, again indicating good agreement. We attribute the overestimation to the high degree of order in the fiber axis we impose in this first effort. Future work will focus on an initial structure with long ladder chains and explore how the initial molecular structure affects the final microstructure and properties of the fibers. The predicted transverse Young's modulus is slightly lower than the experimental values, this is also explained by the perfect nature of our models; $sp^3$ bonds, misaligned crystallites and amorphous regions are expected to significantly increase the stiffness.

The model presented here is a key first step towards predictive computational tools for carbon fibers. Accurate atomic models of microstructures are key for property predictions, not just elastic constants but also ultimate strength. Such predictive tools have the potential to contribute to the design of new carbon fibers with tailored properties.

## Acknowledgements

This work was sponsored by the Institute for Advanced Composites Manufacturing Innovation – The Composites Institute under the Design Modeling and Simulation Technology Area at Purdue University. Support from the Boeing Co and computational resources from nanoHUB and Purdue University are gratefully acknowledged.